\documentclass[twocolumn,showpacs,preprintnumbers,amsmath,amssymb,pra,floatfixm]{revtex4}  %QQ: ADDED FLOATFIX
\usepackage{graphicx}% Include figure files
\usepackage{dcolumn}% Align table columns on decimal point
\usepackage{bm}% bold math
\usepackage{color}

\newcommand{\rb}{$^{87}$Rb~}
\newcommand{\fceo}{$f_{0}$~}
\newcommand{\fr}{$f_{r}$~}
\newcommand{\sh}{S$_{1/2}$}
\newcommand{\ph}{5P$_{1/2}$~}
\newcommand{\pt}{5P$_{3/2}$~}
\newcommand{\ra}{$\rightarrow$~}
\newcommand{\done}{$D_1$~}
\newcommand{\dtwo}{$D_2$~}
\begin{document}
\preprint{7S/5P-Marian}
\title{Direct frequency comb measurements of \\
absolute optical frequencies and population transfer dynamics}

\author{Adela Marian}
\author{Matthew C. Stowe}
\author{Daniel Felinto*}
%\thanks{Present address: Norman Bridge Laboratory of
%Physics, 12-33, Caltech, Pasadena, CA 91125}
\author{Jun Ye}

\affiliation{JILA, National Institute of Standards and Technology
and University of Colorado \\
Department of Physics, University of Colorado, Boulder, Colorado
80309-0440, USA}

\date{\today}

\begin{abstract}

A phase-stabilized femtosecond laser comb is directly used for
high-resolution spectroscopy and absolute optical frequency
measurements of one- and two-photon transitions in laser-cooled \rb
atoms. Absolute atomic transition frequencies, such as the
5S$_{1/2}$ F=2 \ra 7S$_{1/2}$ F"=2 two-photon resonance measured at
788 794 768 921(44) kHz, are determined without \textit{a priori}
knowledge about their values. Detailed dynamics of population
transfer driven by a sequence of pulses are uncovered and taken into
account for the measurement of the 5P states via resonantly enhanced
two-photon transitions.
\end{abstract}

 %Atom manipulation (scanning probe microscopy, laser
%cooling, etc.)
%42.65.Re
%Ultrafast processes; optical pulse generation and pulse compression}

\maketitle

Phase-stabilized optical frequency combs based on mode-locked
femtosecond lasers have formed a powerful connection between the
fields of precision measurement and ultrafast science
\cite{udem99,cundiff-ye03, ye-jstqe}. Numerous applications have
ensued, including measurements of absolute optical frequencies
\cite{Diddams00,Peik04,Gill04} and the development of optical atomic
clocks \cite{Diddams-Science,ye-prl01,nov-Science}. Recent work has
demonstrated that optical frequency combs are a highly efficient
tool for precise studies of atomic structure
\cite{yoon00,Rb-Science}. Direct frequency comb spectroscopy (DFCS)
has been performed on the 5S-5D two-photon transitions in Rb,
permitting high-resolution spectroscopy of all atomic transitions
covered by the comb bandwidth. Additionally, this approach provides
significant advantages for precise studies of time domain dynamics,
coherent accumulation and interference, and quantum control
\cite{Rb-Science}. An extension of frequency comb metrology to the
deep-ultraviolet spectral region (where continuous wave lasers are
not readily available) was recently achieved by using the 4th
harmonic of a femtosecond laser \cite{witte05}.

In this work, we apply DFCS to determine absolute atomic transition
frequencies anywhere within the comb bandwidth, for one- and
two-photon processes. By measuring the previously unknown absolute
frequency of the 5S-7S two-photon transitions in $^{87}$Rb, we show
that prior knowledge of atomic transition frequencies is not
essential for this technique to work, and indicate that it can be
applied in a broad context. When resonant enhancement is enabled by
a comb component tuned near an intermediate 5P state, we observe
two-photon transitions occurring between initial and final states
that differ by one unit of the total angular momentum ($\Delta$F =
$\pm$1), which are absent for far-detuned intermediate states. This
capability of accessing adjacent excited hyperfine levels from the
same ground state allows for direct measurements of hyperfine
splittings. Additionally, we demonstrate that DFCS can be equally
well applied to measuring single-photon transitions and have chosen
the 5S-5P transitions in \rb for such a demonstration. The
measurement of 5P states has also been carried out indirectly via
the 5S-5D two-photon transitions by studying their resonant
enhancement when comb components are scanned through the
intermediate 5P states. We compare the 5P measurements obtained via
one-photon and two-photon DFCS and clearly demonstrate the
importance of population transfer in working with multilevel systems
probed by multiple comb components.

\begin{figure}
\includegraphics[scale=0.36]{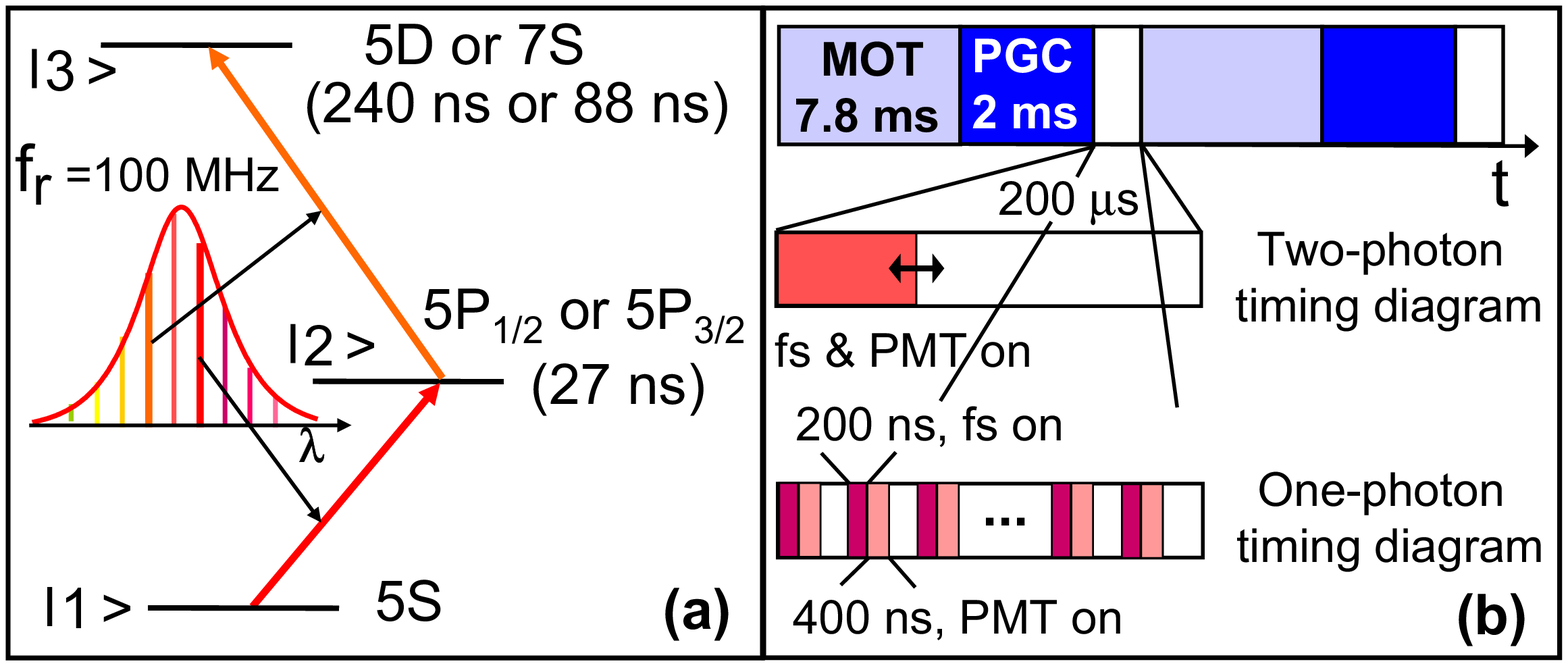}
\caption{\label{Fig1}(Color online) (a) Schematic of the \rb energy
levels participating in the 5S-7S two-photon transitions and 5S-5P
single-photon transitions.  For the two-photon studies, the measured
5S-7S transitions were resonantly enhanced by the 5P states shown.
For the 5P state measurements, the 5S-5P transitions are studied
both directly and indirectly, using the 5P enhancement of the 5S-5D
two-photon transitions. (b) Timing scheme for the 100 Hz experiment
cycle where the 200 $\mu$s zoom window shows the different sequences
for the one- and two-photon measurements. All MOT-related fields are
turned off while probing with the femtosecond laser.}
\end{figure}

The spectrum of a mode-locked femtosecond laser has a set of
discrete optical frequencies that are described by the simple
relation  $\nu_N = Nf_{r}+ f_{0}$, where $N$ is a large integer on
the order of $10^6$. The lines are spaced by $f_{r}$, the pulse
repetition rate, and have a common offset $f_{0}$, the
carrier-envelope offset frequency.  For spectroscopic studies with a
femtosecond laser, both parameters \fr and \fceo are precisely
controlled and stabilized to low-noise optical or radio frequency
oscillators \cite{yecundiff05}.  In the current experiment, the
output of a 20-fs, 100-MHz repetition rate Ti:Sapphire laser with a
full width at half maximum of $\sim$55 nm is used to directly
interrogate a sample of laser-cooled \rb atoms. For the two-photon
case, we study the transitions from the ground 5{\sh} state to the
excited 7{\sh} state, as shown in Fig.~\ref{Fig1}(a). The two
independent parameters \fr and \fceo provide freedom in choosing an
appropriate frequency comb for control of resonant signal
enhancement via the intermediate \ph and \pt states.  The relevant
lifetimes are 88 ns for the 7S states \cite{orozco04} and 27 ns for
the 5P states.  The excited state population is determined from the
7S-6P-5S radiative cascade: the 7S atoms decay to the 6P state and
then down to the ground state, emitting photons at 420 nm. These
blue photons are detected with a photomultiplier tube (PMT) centered
at 420 nm and are counted and subsequently time-binned with a
multi-channel photon counter. The typical loading cycle used for the
magneto-optical trap (MOT) is 100 Hz and the sequence of the
experiment is as follows [Fig.~\ref{Fig1}(b)]:  the atoms are loaded
in the MOT for 7.8 ms, then the quadrupole magnetic field for the
trap is switched off, the atoms are cooled with polarization
gradient (PGC) for 2 ms, then all the MOT beams are extinguished,
the femtosecond comb beam is switched on for 200 $\mu$s using a
Pockels cell (8-ns rise time), and finally, the 420 nm fluorescence
is detected.

For the one-photon studies we have investigated both the \done and
\dtwo transitions in $^{87}$Rb, namely transitions from the ground
state to the \ph excited state at 795 nm and to the \pt state at 780
nm [Fig.~\ref{Fig1}(a)]. We directly detect the fluorescence from
the two 5P states with a near-infrared PMT coupled with a 3-nm
bandwidth interference filter centered at the appropriate wavelength
for the transition. Background counts are further minimized by
spatial filtering and photon collection during the probe laser-on
period is disabled by switching off the PMT. As can be seen in the
one-photon timing diagram of Fig.~\ref{Fig1}(b), during the 200
$\mu$s probe window, we have a sequence of short cycles with the
probe laser on (200 ns) followed by the PMT on (400 ns) to detect
photons from the fast-decaying 5P states. A 2.6 $\mu$s interval (PMT
switch-off time) is required before initiating the next laser cycle.
Both the two-photon and the single-photon signals are averaged over
hundreds of 10 ms experiment cycles.

To null the stray magnetic fields, we apply bias-coil compensation
in three orthogonal directions and make use of the two-photon signal
itself: Zeeman-shifted spectra are obtained with right and left
circularly polarized femtosecond comb light along each direction and
the zero-frequency shift point is determined within $\pm$ 20 kHz.
After nulling the residual B field, there remain two dominant
sources of systematic frequency shifts, both associated with the
femtosecond laser. The first is the radiation pressure of the probe
laser on the atoms and the second is the AC Stark shift \cite
{Rb-Science}. For the present two-photon experiment, the line-center
values are extrapolated to zero interrogation time to suppress
shifts associated with photon-momentum transfer and the transitions
are probed on optimal resonance (i.e., zero detuning for both the
intermediate state and the final state) to minimize the AC Stark
shifts.

\begin{figure}
\includegraphics[scale=0.345]{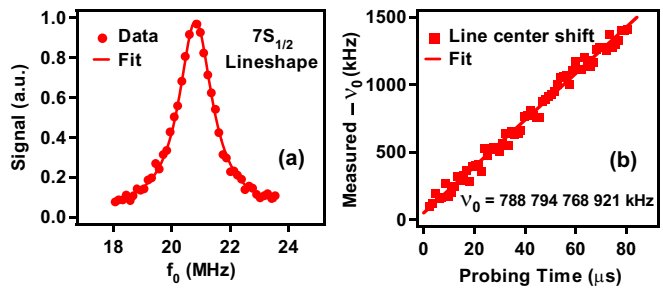}
\caption{\label{Fig2}(Color online) (a) Typical 5{\sh} F=2 \ra
7{\sh} F"=2 two-photon Lorentzian lineshape obtained from a scan of
the offset frequency \fceo for a fixed value of the repetition rate
\fr. (b) Frequency shift of the transition line center vs. probing
time resulting from the momentum transferred by the femtosecond
laser to the cold \rb atoms.  Extrapolation to zero interrogation
time determines absolute atomic transition frequencies free of
radiation pressure effects.}
\end{figure}

A theoretical model describing the interaction of the femtosecond
comb with atoms accounts for detailed dynamics of population
transfer among the atomic states involved in transitions within the
comb bandwidth. The density matrix for the state of the atomic
system is calculated starting with the Liouville equation, with
radiative relaxation included via phenomenological decay terms.
Impulsive optical excitation followed by free evolution and decay is
used to model the interaction with each pulse in the train. The
density matrix equations are solved to a fourth order perturbative
expansion in the electric field and an iterative numerical scheme is
employed to obtain the state of the atomic system after an arbitrary
number of pulses \cite{felinto03, Rb-Science}. This model is applied
to accurately predict the coherent population accumulation in the
relatively long-lived 5D or 7S states, followed by incoherent
optical pumping. Especially important for the indirect 5P
measurements is the incoherent optical pumping to the ground state
hyperfine levels, which depends critically on the 5P state detunings
and will be discussed in detail below.

We begin by discussing the 5S-7S two-photon frequency measurements.
Shown in Fig.~\ref{Fig2}(a) is a typical 7{\sh} F"=2 Lorentzian
lineshape, generated by stepping the offset frequency \fceo for a
fixed value of the repetition rate \fr and recording the subsequent
blue fluorescence corresponding to the coherently accumulated 7S
population. For each data point, to obtain a better signal-to-noise
ratio, the 80 ns-binned counts arising from the fluorescence are
integrated over 2.4 $\mu$s. Alternatively, this lineshape is
retrieved by sweeping \fr with \fceo fixed at some convenient value.
In general, sweeping \fr has the advantage of yielding all the
transitions within the laser bandwidth in only a $\sim$26 Hz scan.
This is due to the fact that the ratio of one-photon optical
transition frequencies (participating in stepwise two-photon
transitions) to \fr is  $\sim$4$\times10^6$, so that for a change in
\fr of $\sim$26 Hz the resonant enhancement is repeated by the next
neighboring comb component. Once the lineshape has been acquired,
what remains is to identify the correct mode number $N$ associated
with each transition. If the optical frequency is already known to
within $f_{r}/2$, this is a straightforward task. For the case of
the 5S-7S two-photon transitions, where the resonance frequencies
are not known \textit{a priori} ($\nu_{opt} = (N_1+N_2)f_{r}+
2f_{0}$), we scan the resonances for two different values of \fr and
unambiguously deduce the sum of the two associated integers ($N_1 +
N_2$) \cite{hansch01, witte05}.  In our case, the two repetition
rates used are separated by 600 kHz to eliminate possible
uncertainties arising from estimations of the \fceo value
corresponding to the peak of the resonance.

After identifying the comb numbers associated with the transition
and reducing the systematic error arising from AC Stark shift, the
remaining error from radiation pressure is suppressed by
extrapolating to zero interrogation time, as shown in
Fig.~\ref{Fig2}(b). We determine for the 5{\sh} F=2 \ra 7{\sh} F"=2
and the 5{\sh} F=1 \ra 7{\sh} F"=1 two-photon transitions in \rb the
absolute optical frequencies of 788 794 768 921(44) kHz and 788 800
964 199 (122) kHz, respectively. The excited state hyperfine
interval of 639 404 (130) kHz agrees very well with a previous
differential measurement performed with a picosecond pulsed laser
\cite{snadden96}. The transition spectra reported in
\cite{snadden96}, as well as a CW-based scan \cite{taiwan04},
indicated that the F=2 \ra F"=2 and the F=1 \ra F"=1, i.e.
$\Delta$F=0 transitions, were the only allowed 5S-7S transitions in
$^{87}$Rb. However, we observe additional lines, as the resonant
intermediate 5P state also enables the F=2 \ra F"=1 and F=1 \ra
F"=2, i.e., $\Delta$F=$\pm 1$, two-photon transitions. Similar
$\Delta$F=$\pm 1~$ S-S transitions have been previously observed in
Na in a two-step excitation experiment employing two tunable CW dye
lasers, which enabled a direct measurement of the excited state
hyperfine splitting \cite{vialle74}.

\begin{figure}
\includegraphics[scale=0.52]{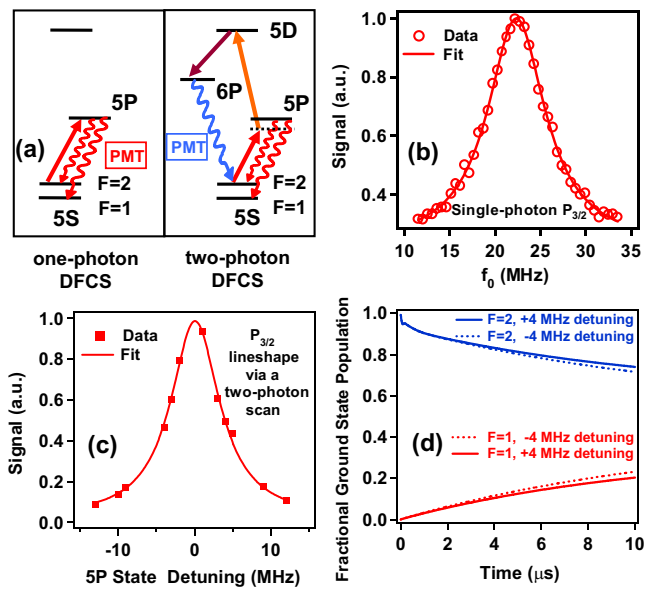}
\caption{\label{Fig3}(Color online) (a) Schematic of one- and
two-photon DFCS, used for measuring single-photon transition
frequencies. (b) Lineshape of the 5{\sh} F=2 \ra \pt F'=3 transition
obtained from a scan of \fceo for a fixed value of $f_{r}$, by
one-photon DFCS. (c) Same lineshape as in (b) retrieved using its
resonant enhancement of the 5{\sh} F=2 \ra \pt F'=3 \ra 5D$_{5/2}$
F"=4 closed two-photon transition, as a function of the detuning
from the intermediate state. (d) Theoretical plot of the time
evolution of the ground state populations for two (symmetric)
detuning values in (c), showing that (i) most of the atoms remain in
the initial F=2 ground level for this closed two-photon transition
and (ii) the ground state populations are largely insensitive to the
sign of the detuning.}
\end{figure}

DFCS also works well for absolute frequency measurements of
single-photon transitions. DFCS results on the 5P state energy
levels are obtained in one-photon and two-photon measurements and
compared. The one-photon DFCS employs radiative detection directly
from the 5P states (Fig.~\ref{Fig3}(a) left panel), while the
two-photon DFCS studies the 5P states indirectly, via resonant
enhancement of the 5S-5D two-photon transitions as a function of the
detuning from the intermediate 5P states (Fig.~\ref{Fig3}(a) right
panel). First, we measure the 5{\sh} F=2 \ra \pt F'=3 \dtwo
transition with one-photon DFCS, with the resultant transition line
shown in Fig.~\ref{Fig3}(b). Frequency scans are carried out
similarly to those of the 5S-7S lines, that is, by stepping \fceo
continuously while keeping \fr fixed. The absolute optical frequency
measured for this transition is 384 228 115 271 (87) kHz. For the
two-photon DFCS, we use a set of different pairs of \fr and \fceo
specifically chosen to have varying detunings from the 5P state for
each data point shown in Fig.~\ref{Fig3}(c), while at the same time
satisfying the 5S-5D two-photon resonance (these transitions have
the same decay channels as the 5S-7S transitions and are also
monitored via the 420 nm fluorescence). The lineshape in
Fig.~\ref{Fig3}(c) is retrieved by detecting the 420 nm signal as a
function of 5P state detuning and the optical frequency measured by
this two-photon DFCS is 384 228 115 309 (63) kHz, in agreement with
the result obtained from one-photon DFCS within the standard
deviation.  It is important to mention that for this scan we take
advantage of the 5{\sh} F=2 \ra \pt F'=3 \ra 5D$_{5/2}$ F"=4 being
the only 5S-5D `closed transition'.  As shown in theory plots in
Fig.~\ref{Fig3}(d), this closed transition ensures that most of the
atoms initially starting out in the F=2 ground-state hyperfine level
remain in that level, while $\sim$20$\%$ of the atoms fall into the
dark F=1 ground state due to optical pumping and hence do not
contribute to the signal. In addition, the probe laser power is
sufficiently reduced for the two-photon DFCS experiments to further
decrease optical pumping effects.

\begin{figure}
\includegraphics[scale=0.52]{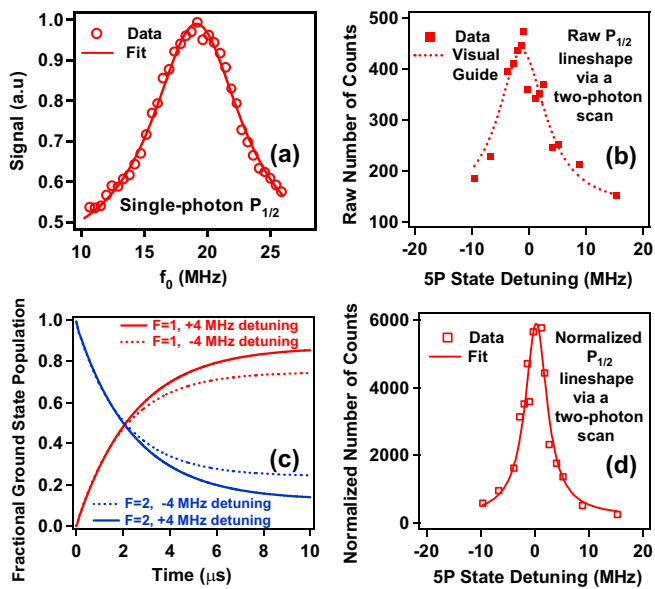}
\caption{\label{Fig4}(Color online) (a) Lineshape of the 5{\sh} F=2
\ra \ph F'=2 transition obtained from a scan of \fceo for a fixed
value of $f_{r}$, by one-photon DFCS. (b) Raw counts for the same
lineshape as in (a) by two-photon DFCS, along with a visual guide
for the data. (c) Theoretical plot of the time evolution of the
ground state populations for two (symmetric) detuning values in (b)
showing a significant difference in the population transfer between
the ground state levels due to optical pumping caused by varying
detunings from the other 5P states. (d) Normalized lineshape
corresponding to the raw data in (b), obtained by using results from
the theory simulation in (c) accounting for optical pumping
effects.}
\end{figure}

Next, we employ DFCS to study another single-photon transition in
the \done manifold, 5{\sh} F=2 \ra \ph F'=2, as shown in
Fig.~\ref{Fig4}(a). Again, \fceo is scanned while \fr is stabilized
to a convenient value. The absolute optical frequency for this
transition is determined to be 377 105 206 563 (184) kHz, in
agreement with a previous wavelength-based measurement \cite{npl91}.
For the corresponding two-photon DFCS experiment we map the 5{\sh}
F=2 \ra \ph F'=2 \ra 5D$_{3/2}$ F"=3 two-photon transition in the
same manner employed for Fig.~\ref{Fig3}(c). Figure~\ref{Fig4}(b)
shows the raw data yielded by these ($f_{r}$, $f_{0}$) pair
selections, along with a visual guide for the data. The apparent
linewidth is significantly broader than that associated with the 5P
state. Unlike the previous two-photon DFCS measurement reported in
Fig. 3, the pairs of \fr and \fceo used to obtain each point in
Fig.~\ref{Fig4}(b) lead to substantially different detunings of the
other 5P states and subsequently, varying optical pumping to the F=1
ground state. Indeed, the theory model applied to the actual
experiment conditions predicts significantly different ground state
population transfer dynamics. As shown in Fig.~\ref{Fig4}(c), the
asymptotic values of the F=2 ground-state population are not the
same for symmetric detunings from the intermediate state.
Figure~\ref{Fig4}(d) presents the Lorentzian lineshape resulting
from the normalization of the raw data shown in Fig.~\ref{Fig4}(b)
with respect to the theoretical value of (1 - $\rho_{F=1}$), where
$\rho_{F=1}$ is the fractional ground state population in F=1, as
shown in Fig.~\ref{Fig4}(c). After implementing this normalization,
the optical frequency for the transition measured by the two-photon
DFCS is 377 105 206 939 (179) kHz, within the error bars of the
corresponding one-photon DFCS result. We note that for all
measurements reported in the paper, the statistical errors (one
standard deviation of the mean) associated with \pt are
significantly smaller than those associated with 5P$_{1/2}$.  This
is due to the stronger transition strength and less severe optical
pumping effects for \pt F=3 (part of a closed transition), leading
to larger signal-to-noise ratios.

In conclusion, a phase-stabilized femtosecond comb has been used as
an effective tool to perform direct spectroscopy of one- and
two-photon transitions in cold \rb atoms. We have demonstrated that
DFCS can be successfully applied to one-photon studies, by measuring
5{\sh} \ra 5P$_{1/2, 3/2}$ transitions both directly and indirectly,
via their resonant enhancement of the 5S-5D two-photon transitions.
Additionally, we have shown the importance of including the dynamic
population changes arising from pulse-accumulated population
transfer in this indirect one-photon measurement. Finally, we have
demonstrated that by using DFCS, the absolute value of the 5{\sh}
\ra 7{\sh} two-photon transitions in \rb is conclusively determined,
with no \textit{a priori} knowledge about their optical frequency.

We thank J. R. Lawall and J. L. Hall for technical help and
discussions. Funding for this research is provided by ONR, NSF,
NASA, and NIST.

*Present address: Norman Bridge Laboratory of Physics, 12-33,
Caltech, Pasadena, CA 91125.
%\end{acknowledgments}

\end{document}